# A Radiation Tolerant Proton Detector Based on MAPbBr$_3$ single crystal

Huaqing Huang[1], Linxin Guo[1], Yunbiao Zhao[1], Xinwei Wang[2], Wenjun Ma[1] and Jianming Xue[*,1,3]

*1State Key Laboratory of Nuclear Physics and Technology, School of Physics, Peking University, Beijing 100871, P. R. China;*

*2School of Advanced Materials, Shenzhen Graduate School, Peking University, Shenzhen 518055, P. R. China;*

*3CAPT, HEDPS and IFSA, College of Engineering, Peking University, Beijing 100871, P. R. China.*

*\*Corresponding author. Email: jmxue@pku.edu.cn*

## Abstract

The performance and radiation tolerance of the proton detector based on MAPbBr$_3$ perovskite single crystal are investigated here with 3MeV protons. The detector can monitor fluence rate and dose quantificationally at a low applied bias electric field(0.01V/μm) within a dose range of 45 kGy. The detector can also be worked at zero bias due to the Dember effect. The dark current of the detector reduced to 20% of the initial value after being irradiated with protons to a total fluence of $7.3 \times 10^{13}$ p/cm$^2$ (1 MGy), however, it can be recovered at room temperature within hours. These results suggest that this kind of detector has a promising application in proton therapy and proton imaging etc.

## Introduction

Proton detection is of great interest in a wide range of applications, such as high energy physics, accelerator beam diagnostics, deep space exploration, especially in proton imaging and proton therapy.[1-6] There are types of detectors for radiation detection, including calorimeters, radiographic films, gas detectors, scintillator detectors and semiconductor detectors. Among these detectors, the semiconductor detectors seem to be appropriate for proton detection in real time due to their excellent energy resolution, rapid time response, small size and system integrability[7, 8].

However, the performance of semiconductor detectors will be degraded under long-term proton irradiation, resulting from the defects like vacancies and interstitials caused by energetic protons. For example, it has been found that the reverse current of silicon detector was increased by an order when irradiated with 3 MeV, $10^{13}$ p/cm$^2$ protons[9]. Zou et al. found that the charge collection efficiency (CCE) of the diamond detector was reduced to 20% of the initial value after being irradiated with 800 MeV, $2\times10^{14}$ p/cm$^2$ protons[10]. Therefore, it is essential to develop a proton detector with high radiation tolerance as the demand for proton detection is increasing in proton therapy and high energy physics experiments etc.[10, 11].

In recent years, metal halide perovskites exhibit potential for radiation detection due to their excellent carrier transport properties, strong stopping power, low costs, and high radiation tolerance [12, 13]. To date, they have already been demonstrated to detect X-rays, gamma rays, alpha particles, electrons, and neutrons [14-18]. Deumel et al. showed that the X-ray induced current of the perovskite detector was stable with a dose of 11 Gy$_{air}$[15]. Wei et al. found that there was no significant degradation of the perovskite detector after gamma ray irradiation at a total dose of 350 Gy[19]. Moreover, Lang et al. discovered that perovskite-based solar cells show superior radiation resistance compared with silicon and silicon carbide-based solar cells under 68 MeV protons irradiation[20, 21]. Therefore, perovskites seem to be a promising material for proton detection with high radiation tolerance.

However, there is no report on perovskite-based proton detector to date. Meanwhile, most of current works on proton irradiation of perovskite are focused on solar cells with complicated structures, while the influences of transport layers and glass cannot be excluded[20-24].

Therefore, it is significant to investigate the performance and radiation tolerance of the detector with a simple structure based on perovskite.

In this work, a planar structure proton detector based on methylammonium lead tribromide (MAPbBr$_3$) single crystal was fabricated and tested. The detector shows a linear response to the fluence rate and absorbed dose of 3 MeV protons within a dose range of 45 kGy. The radiation tolerance of the detector is studied by recording the dark current in situ under proton irradiation with an accumulated fluence of $7.3 \times 10^{13}$ p/cm$^2$, i.e. a total dose of 1MGy. The dark current reduces to 20% of the initial value at the end of irradiation, however, it can be recovered at room temperature within hours. The mechanism of degradation and self-healing behaviors are discussed at last. This study indicates that the halide perovskite is a promising proton detector candidate, which will promote the further investigation.

## Results and discussion

**Direct proton detection.** The MAPbBr$_3$ single crystals were grown by inverse temperature crystallization method[25]. Figure 1a is the XRD spectrum of the as-grown crystal, and the four sharp peaks are related to (100), (200), (300) and (400) lattice planes of cubic MAPbBr$_3$ perovskite, indicating that the single crystal is of high quality[26, 27]. The inset in Figure 1a shows one of single crystals with a dimension of $6 \times 6$ mm.

For demonstration of the proton detection, a planar detector with Au/Ti/MAPbBr$_3$/Ag structure was fabricated as shown in Figure 1b, and the active volume is about $2 \times 1.5 \times 1$ mm. The performance of the detector was tested with an in-situ radiation test system based on 4.5 MV electrostatic accelerator [28]. Proton with energy of 3 MeV was used here and the beam current was from 10 nA to 85 nA (the corresponding fluence rate was from $2.65 \times 10^{10}$ cm$^{-2}$s$^{-1}$ to $2.26 \times 10^{11}$ cm$^{-2}$s$^{-1}$). Generally, electrons and holes will be generated along the proton track due to the ionization. Figure 1c depicts the electronic energy loss of 3 MeV protons in MAPbBr$_3$ calculated with the Stopping and Range of Ions in Matter (SRIM) code [29]. The Bragg peak is located at 97 μm, which means a large number of carries will be generated in the detector. And it also means that the detector is capable of detecting protons with higher energy.

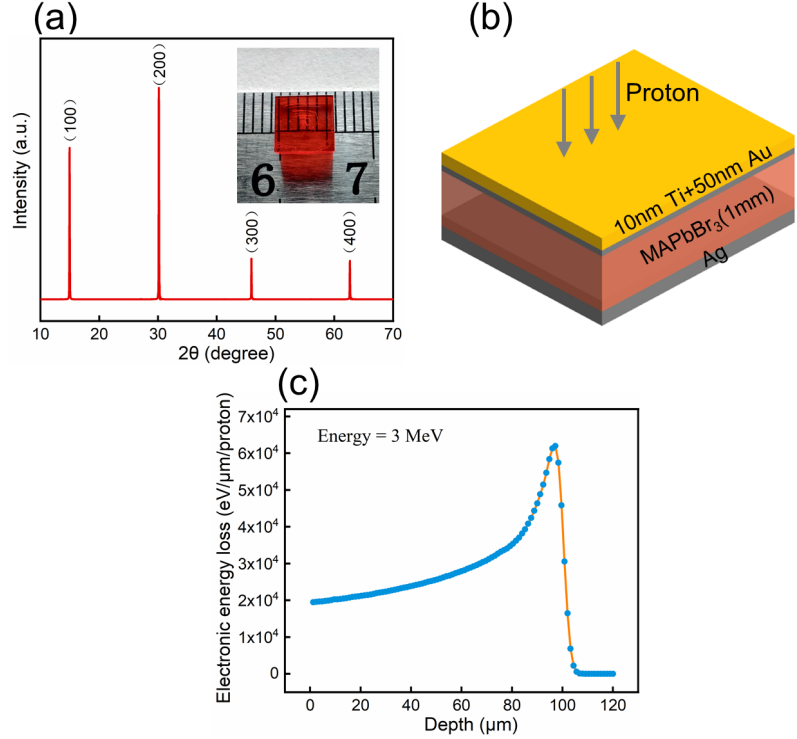

**Figure 1. Device Characterization and fabrication.** (a) XRD patters of the as-grown single crystal. The inset is a photograph. (b) Schematic architecture of the MAPbBr₃ proton detector. (c) The depth distribution of electronic energy loss of 3 MeV protons in MAPbBr₃ calculated with SRIM.

Figure 2e-2f show the results of 3 MeV, $2.65 \times 10^{10}$ cm$^{-2}$s$^{-1}$ protons detection. The dark current and a typical current-voltage (I-V) response under proton irradiation is shown in Figure 2a. The dark I-V is symmetric without rectification effect, which indicates the device is a resistor-type radiation detector with ohmic contact. However, the proton induced current under negative bias is larger than that under positive bias. Therefore, the detector was worked under negative bias for a greater response. Figure 2b shows a time-resolved current response at an applied electric field of $10^{-4}$ V/μm, i.e. a bias voltage of -0.1 V. The proton current rises and falls rapidly without charges accumulation when the proton beam is switched on and off, which demonstrate the real-time detection capability of the detector.

The proton induced charges collected by the detector during a period of time can be obtained by integrating the current. The collected charges increase linearly with the increasing fluence as shown in Figure 2c at different bias voltage. The slope of the linear fit is defined as the sensitivity of the detector. Figure 2d shows the positive dependence of sensitivity on the value of applied bias voltage. Although a higher sensitivity can be obtained by increasing the bias further, the detector was worked at -10 V (0.01V/μm) to avoid the notorious ion migration in perovskite, and moreover, a low bias is meaningful for applications in space etc[30, 31].

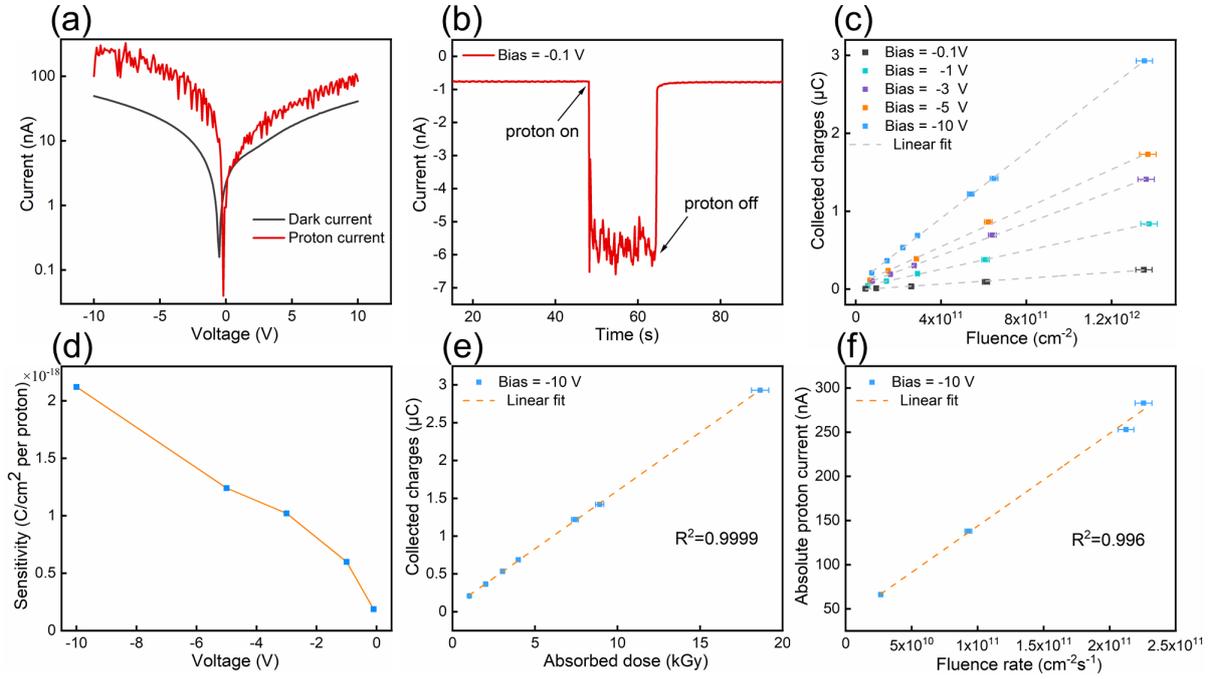

**Figure 2. 3MeV protons detection by MAPbBr₃ detector.** (a) Dark current and proton induced current of the detector. (b) Time-resolved proton induced current at -0.1V bias (10⁻⁴V/μm). (c) The relationship between the collected charges and the fluence at different bias voltage. (d) The sensitivity of the detector at different bias voltage. (e) The linear relationship between the collected charges and the absorbed dose. (f) The linear current response to the fluence rate.

The dose monitoring is essential in proton therapy and other radiation environments[32, 33]. The dose absorbed by the detector can be calculated by

$$D = \frac{E \times F}{\rho \times d} \ (Gy) \tag{1}$$

where $E$ is the energy the detector absorbed of a proton, $F$ is the fluence, $\rho$ is the density of MAPbBr₃ (3.83 g/cm³), $d$ is the projected range of protons. Figure 2e shows a perfect linear relationship between the collected charges and the absorbed dose at a bias voltage of -10 V, which indicates that the detector can monitor the dose accurately and quantificationally. Moreover, the sensitivity of the detector can be as high as (1.540 ± 0.008)×10⁻¹⁰ C/Gy, which is 30 times higher than that of the organic thin-film proton detector[34]. In addition, it is worth noting that the total dose of Figure 2e reaches 45 kGy, which is equivalent to the dose of 1 billion times proton radiography, or the dose absorbed inside the spacecraft for 0.4 million years[35, 36]. Therefore, this detector will maintain its excellent performance when applied to these applications.

The fluence rate detection in real-time is in demand in space and accelerators[37]. Figure 2f displays the linear response of the current value to the fluence rate of protons, which means the detector can monitor the fluence rate in real time. The detection limit is about $2.9 \times 10^9$ cm$^{-2}$s$^{-1}$, obtained by extrapolating the current to three times of the root mean square of the dark current. It should be expected that a lower detection limit can be obtained by suppressing the dark current, for example, using MAPbCl$_3$ perovskite with lager band gap, doping to compensate or using a guard ring electrode [19, 38].

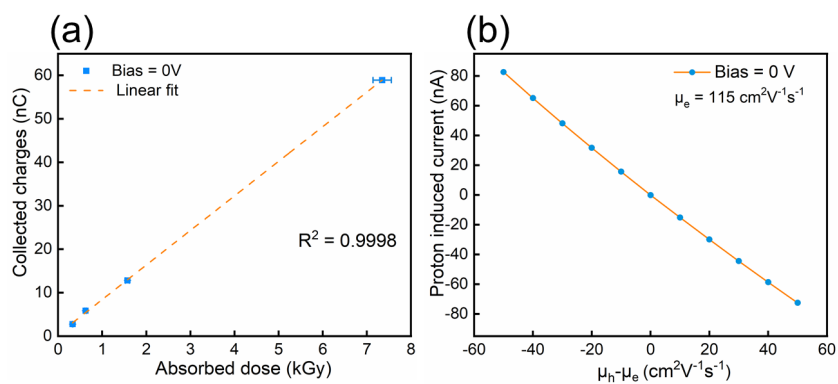

**Figure 3. Experiment and simulation results of proton detection at zero bias. Experiment result**: (a) The induced charges that the detector collected of 3MeV protons at zero bias versus the absorbed dose. **Simulation result**: (b) The influence of electron and hole mobility on the proton induced current at zero bias. The current is zero when $\mu_h = \mu_e$.

Self-powered radiation detectors are of great importance in energy-sparse and harsh environments such as in space[39]. As mentioned in Figure 2a, the detector is ohmic contact without internal potential barrier, however, it represents a significant current response under proton irradiation at zero bias, and the sensitivity is $(7.96 \pm 0.07) \times 10^{-12}$ C/Gy as shown in Figure 3a. It is mainly caused by the Dember effect due to the different mobility of electrons and holes. Specifically, a plenty of electron-hole pairs will be produced localized around the Bragg peak under proton irradiation and a carrier concentration gradient will be formed. As electrons diffuse faster than holes generally, a net diffusion current will be generated, and resulting a Dember electric field along the incidence direction of protons[40, 41].

In order to confirm this hypothesis, a simulation was performed with COMSOL

Multiphysics software. The parameters used in simulation are listed in Table S1[42-44]. It should be noted that some parameters such as hole mobility and lifetime are assumed values because the data measured in literatures vary widely[43]. However, the actual values are not important since the simulation is mainly focused on studying the influence of parameters. The carrier generation rate can be calculated by:

$$G_n(x) = G_p(x) = \frac{S_e(x) \times f}{\omega} \ (\text{cm}^{-3}\text{s}^{-1}) \tag{2}$$

where $G_n(x)$ and $G_p(x)$ is electron and hole generation rate respectively, $S_e$ is the electronic energy loss shown in Figure 1c, $f$ is the fluence rate of proton, $x$ is the depth, and $w$ is the average ionization energy of MAPbBr₃. Additional details about simulation can be found in Methods.

Figure 3b shows the influence of hole mobility on current at zero bias under proton irradiation when the electron mobility is fixed at 115 cm²V⁻¹s⁻¹. The current is zero when their mobility is equal, however, there is a significant current response when the mobility is different. Actually, Cao et al. fabricated a self-driving photodetector based on MAPbI₃ thin film via Dember effect recently[40]. The result of this study shows that there also exits Dember effect under proton irradiation in MAPbBr₃. The perovskite-based detector with ohmic contact here provides a new perspective for self-powered proton detection.

**performance degradation.** The radiation tolerance is an important figure of merit for a proton detector. Generally, the incident energetic protons transfer energy to the target material through nonionizing or ionizing energy loss (NIEL and IEL) mechanisms. NIEL leads to atoms displacements, generating vacancies and interstitials defects, which will act as recombination or trapping centers and significantly influence the performance of the detector[45]. Therefore, the 3 MeV proton is appropriate to investigate the radiation tolerance of the detector since the defects are mainly around the Bragg peak, which is inside the detector as shown in Figure S1.

The proton induced current remains stable without significant degradation as shown in Figure 2b with an accumulated fluence of 5.4×10¹¹cm⁻². This result is constituent with the work of Lang et al., in which they found that the proton induced quantum efficiency of the perovskite-based solar cell degraded by only 7% at a fluence of 10¹² cm⁻² under 68 MeV protons irradiation, while the SiC diode decreased by 75% under the same condition[21]. It should be noted that the displacement defects caused by 68 MeV protons is much less than that of 3 MeV protons. Meanwhile, the detector maintains a linear response to the absorbed dose within a total dose of 45 kGy as shown in Figure 2c. Hence, the detector is sufficient radiation tolerant in practical applications.

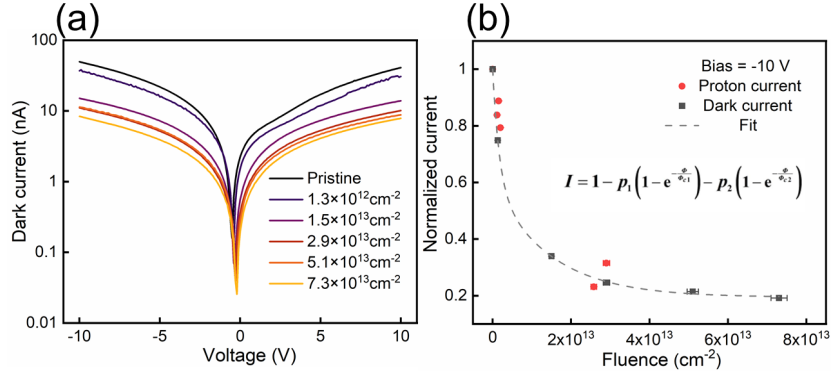

**Figure 4. The electric properties variation under long-term proton irradiation.** (a) Dark I-V curves of the detector before irradiation and at different proton fluence after irradiation recorded in situ. (b) Normalized dark current (dark dots) and proton current (red dots) of the detector at -10V bias as a function of the proton fluence, the dashed line is the fitting curve of the dark current.

In order to study the radiation tolerance of the detector in harsh radiation environment, it was irradiated to a total fluence and dose of $7.3 \times 10^{13}$ cm$^{-2}$, 1 MGy respectively. And the dark current-voltage curves were recorded in situ at different fluence as shown in Figure 4a. Apparently, the dark current decreases gradually with the increase of fluence. This is mainly due to the charge collected efficiency degradation under long-term proton irradiation, which also occurs in other semiconductor detectors[46-48]. Therefore, although the collected charge remains linear with the absorbed dose at the end of irradiation as shown in Figure S2, the sensitivity decreases to $(4.4 \pm 0.1) \times 10^{-11}$ C/Gy. In general, proton irradiation can create displacement damages such as vacancies and interstitials, which will act as generation and recombination centers of carriers, leading to minority carrier lifetime reduction and removal effect of carriers, and eventually, resulting in the decrease of the dark current[46, 47].

The decay of the dark current value with fluence can be fitted with a double exponential function as shown in Figure 4b, which has been used to fit proton current degradation of silicon carbide and diamond detector [46, 47]. It can be seen that the decay trend of proton induced current is constituent with the dark current. The dark current remains 20% of the initial value at the end of irradiation with an accumulated fluence of $7.3 \times 10^{13}$ cm$^{-2}$. Similarly, the diamond detector, one of the most radiation resistance detectors, remains about 20% of the initial proton current when irradiated with $2 \times 10^{14}$ cm$^{-2}$ protons [10]. However, the NIEL of 800MeV protons in diamond is two orders of magnitude lower than that of 3MeV protons in MAPbBr$_3$, which means much less displacements damages. Moreover, the higher IEL of 800 MeV protons can result in defects annealing during irradiation[45]. Hence, it can be speculated that the MAPbBr$_3$ detector has a radiation tolerance which is comparable to the diamond detector.

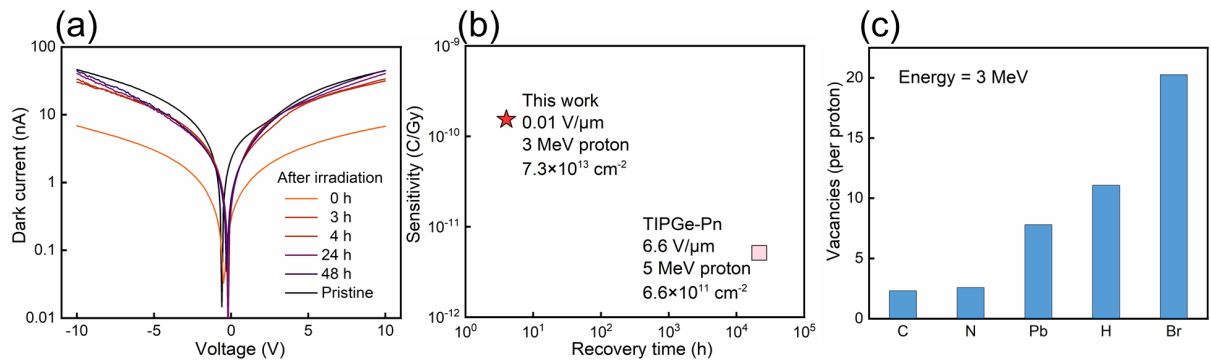

**Figure 5. The self-healing behavior of the detector.** (a) Dark I-V curve recovery of the detector with time at room temperature after irradiation. (b)A performance comparison of the MAPbBr$_3$ detector in this work and the organic thin-film proton detector in Ref.34. (c) Vacancies of each element caused by 3 MeV protons calculated with SRIM.

**Self-healing.**    After irradiation, the detector was placed in air at room temperature and the dark I-V curves were measured with increasing elapsed time. As shown in Figure 5a, the dark current quickly recover to 80% of the initial value within three hours right after irradiation, and then recover to approach the initial state slowly in several days. Generally, defects caused by irradiation can be healed under thermal annealing at high temperature. For example, the dark current of the Silicon and Thallium bromide detector can be recovered close to the initial value by high-temperature thermal annealing after protons irradiation[9, 49]. Recently, Fratelli et al. demonstrated a proton detector based on organic thin film with a sensitivity of $(5.15 \pm 0.13) \times 10^{-12}$ C/Gy, and its dark current was recovered to the initial value after three months when irradiated with 5MeV, $6.6 \times 10^{11}$ cm$^{-2}$ protons [34]. As compared in Figure 5b, the MAPbBr$_3$ detector in this work has a speedy self-healing characteristic and high sensitivity, which makes it promising for applications which requires intermittent irradiation, such as proton therapy, proton imaging and beam diagnostics etc.

It has been proved that the phase structure and energy band of perovskite films almost unchanged under protons irradiation at a fluence of $10^{14}$ cm$^{-2}$[50]. In order to figure out the mechanism of the degradation and self-healing characteristics of the detector, the total vacancies of each element in MAPbBr$_3$ caused by 3 MeV protons is calculated with SRIM. As can be seen in Figure 5c, Br vacancies accounted for the largest proportion. The first-principles calculation indicates that the formation energy and migration barrier are very low for both Br vacancies ($V_{Br}$) and interstitials ($I_{Br}$) in MAPbBr$_3$, which means that they can migrate at room temperature while other defects are stable[51, 52]. Therefore, combined with Figure 5a, it can be speculated that it is defects related to Br which dominate the performance degradation, and then, these defects can migrate to the original lattice positions at room temperature, resulting in the self-healing after irradiation. This self-healing behavior further improves the radiation tolerance of the detector.

## Conclusions

In summary, a proton detector based on metal halide perovskite was fabricated and tested in this study. The detector can monitor proton dose quantificationally up to a total dose of 45 kGy with a high sensitivity of $(1.540 \pm 0.008) \times 10^{-10}$ C/Gy. And it has a current response at zero bias, which is mainly due to the Dember effect according to the simulation. Although the dark current of the detector reduces to 20% of the initial value under 3 MeV, $7.3 \times 10^{13}$ cm$^{-2}$ (1 MGy) protons irradiation, the radiation tolerance of the detector is comparable to the diamond detector. Most importantly, the dark current can be recovered in several hours at room temperature, which further improves the radiation tolerance of the detector. The degradation and self-healing behaviors are mainly due to the generation and migration of defects related to Br. In a word, this study demonstrates the immense potential of metal halide perovskites for proton detectors with high sensitivity, high radiation tolerance, self-driving and self-healing characteristics. These results suggest that the metal halide perovskite is a promising candidate for proton detection in proton therapy, proton imaging, beam diagnostics and space etc.

## Experimental Section

**Synthesis of MAPbBr$_3$ single crystals**. The MAPbBr3 single crystals were grown using inverse temperature crystallization (ITC) method previously reported [25]. Methylammonium bromide (MABr) and lead bromide (PbBr2) were dissolved at a 1:1 molar ratio in dimethylformamide (DMF) solvent to prepare a solution of 1.6 M. The solution was stirred at room temperature for 3-5h, followed by filtration to obtain the precursor. Then, 3 mL precursor solution was transferred into a vial and placed in a vacuum drying oven with rising temperature to 60 ℃ slowly at a ratio of 10℃/h. Several crystals can be obtained vary from 2 to 6mm after 24-48h at 60 ℃. Finally, the crystals were washed with 60 ℃ DMF solution, and dried in vacuum oven at 60 ℃ for 1-2h.

**Device fabrication**. E-beam evaporation method was used to deposit the metal electrode (10nmTi and 50nm Au) on one side of MAPbBr$_3$ single crystal, and silver paste was spun on the opposite side. The silver paste was used to lead aluminum wire from electrode to printed circuit board (PCB).

**Measurements and characterization.** The crystal structures of single crystals were characterized by an X-ray diffractometer (PANalytical, Empyrean) equipped with Cu Kα X-ray (λ = 1.5406 Å). The operated voltage and current were 45 kV, 40 mA. The test mode was θ-2θ linkage mode, the scanning range was 10−70°, and the step size was 0.01°. The electrical measurements were performed by a Keithley 4200A-SCS semiconductor characterization system.

**Proton irradiation.** The proton detection and irradiation experiments were performed on in-situ irradiation platform based on 4.5 MV electrostatic accelerator at Peking University of State Key Laboratory of Nuclear Physics and Technology. The energy of protons is 1-3MeV and the proton beam intensity is 10-85nA, the corresponding fluence rate to 10nA beam is $2.65 \times 10^{10}$ cm$^{-2}$s$^{-1}$. The detector was placed in a vacuum chamber of the beam terminal through a PCB interface, and the response was recorded by a Keithley 4200A-SCS semiconductor characterization system remotely. The beam intensity (fluence) was measured by a Faraday cup.

**Calculations and simulation**. The depth distribution of proton atoms and electronic energy loss, and total vacancies were calculated by Stopping and Range of Ions in Matter (SRIM) code using the "full damage cascade" calculation mode. Densities used were Au = 19.31 $g/cm^3$, $MAPbBr_3$ = 3.83$g/cm^3$. Displacement energies used were C = 28 eV, H = 10 eV, N = 28 eV, Pb = 25 eV, Br = 25 eV[45]. The simulation of the detector was performed with COMSOL Multiphysics software (COMSOL, Inc.). The detector was simplified to a one-dimensional model whose length was 1 mm and electrode area was 3 $mm^2$ with ohmic contact. The carriers generate rate was estimated with equation (2). The parameters used in simulation are listed in Table S1.

## Supporting Information Available

Supporting Information is available from the author.

## Acknowledgements

This work is supported by the National Natural Science Foundation of China (Grant No. 2135002), and the Science Challenge Project (No. TZ2018004). The authors are grateful for the computing resources provided by the High Performance Computing Platform of the Center for Life Science of Peking University, and the Weiming No. 1 and Life Science No. 1 High Performance Computing Platform at Peking University.

## Conflict of Interest

The authors declare no conflict of interest.

## Data Availability

The data that support the findings of this study are available from the corresponding author upon reasonable request.

***Supporting Information***

# A Radiation Tolerant Proton Detector Based on MAPbBr₃ single crystal

Huaqing Huang[1], Linxin Guo[1], Yunbiao Zhao[1], Xinwei Wang[2], Wenjun Ma[1]

and Jianming Xue[*,1,3]

*[1]State Key Laboratory of Nuclear Physics and Technology, School of Physics, Peking University, Beijing 100871, P. R. China;*

*[2]School of Advanced Materials, Shenzhen Graduate School, Peking University, Shenzhen 518055, P. R. China.;*

*[3]CAPT, HEDPS and IFSA, College of Engineering, Peking University, Beijing 100871, P. R. China.*

*\*Corresponding author. Email: jmxue@pku.edu.cn*

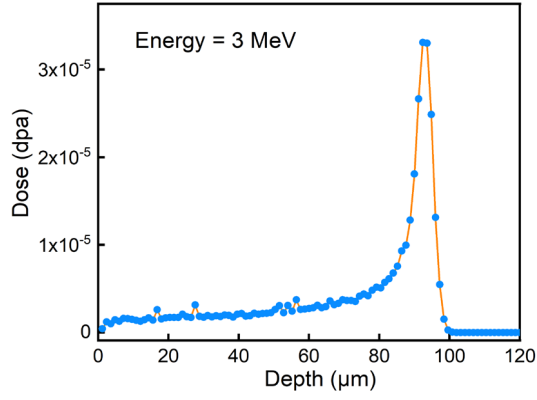

**Figure S1.** The damage distribution of 3 MeV protons in MAPbBr₃ calculated with SRIM.

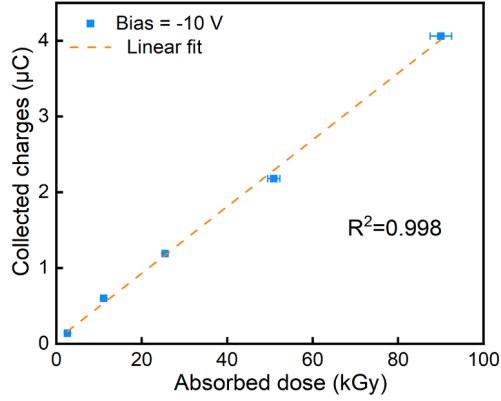

**Figure S2.** The relationship between the collected charges and the absorbed dose at the end of the irradiation with a cumulative fluence of $7.3 \times 10^{13}$ cm⁻². The sensitivity is $(4.4\pm0.1) \times 10^{-11}$ C/Gy.

**Table S1.** Material parameters of the MAPbBr₃ used in simulation [1-3].

| | $E_g$ (eV) | $\mu_e$ (cm²V⁻¹s⁻¹) | $m_n^*/m_p^*(m_0)$ | $\chi$(eV) | $\varepsilon_r$ | $N_A$(cm⁻³) | $\tau$(μs) | $\omega$(eV) |
|---|---|---|---|---|---|---|---|---|
| MAPbBr₃ | 2.3 | 115 | 0.21/0.23 | 3.73 | 4.8 | $10^{10}$ | 10 | 6 |

The parameters of the MAPbBr₃ used in simulation are listed in Table S1, where $E_g$ is the bandgap, $\mu_e$ is the electron mobility, $m_n^*(m_p^*)$ is the effective mass of electron (hole), $\chi$ is the electron affinity, $\varepsilon_r$ is the relative permittivity, $N_A$ is the accepter concentration, $\tau$ is the carrier lifetime assumed for Shockley-Read-Hall recombination, and $\omega$ is the average ionization energy. The hole mobility is from 65 to 165 cm²V⁻¹s⁻¹. It should be noted that many parameters vary a lot in literatures, therefore, the influence of parameters instead of the actual value is studied in simulation.